\documentclass[12pt]{article}

\begin{document}
\title{Stochastic Schr\"odinger evolution 
and symmetric K\"ahler manifolds of low dimension}
\author{L. P. Hughston\\Department of Mathematics\\King's College
London\\The Strand\\London WC2R 2LS\\{\em e-mail:
lane.hughston@kcl.ac.uk}
\and
 K. P. Tod\\The Mathematical Institute\\St Giles'\\Oxford OX1 3LB\\
{\em e-mail: tod@maths.ox.ac.uk} }
\date{}
\maketitle

\begin{abstract}
We consider the manifold-valued, stochastic extension of the 
Schr\"odinger equation
introduced by Hughston
\cite{h1} in a manifestly covariant, 
differential-geometric framework, and examine the
resulting quantum evolution
on some specific examples of K\"ahler manifolds with many symmetries. We
find  conditions on the curvature for the 
evolution to be a `collapse process' in the sense of Brody and 
Hughston \cite{bh1} or, more generally, 
a `reduction process', and give examples that
satisfy these conditions. For some of these examples, we show that the
L\"uders projection postulate admits a consistent interpretation
and remains valid in the nonlinear regime.
\end{abstract}
\section{Introduction}

In reference \cite{h1}, in the context of a geometrical description of quantum 
mechanics, the first author investigated a generalisation of the Schr\"odinger 
equation called the stochastic Schr\"odinger evolution. This 
evolution is governed by a stochastic differential equation (SDE) for a random process
with values in the state-manifold ${\bf CP}^n$, 
the space of one-dimensional subspaces of the usual linear space of 
states ${\bf C}^{n+1}$. The evolution requires for its definition 
various specific differential-geometric features of the state-manifold 
including its standard 
K\"ahler metric, the Fubini-Study metric. 

In reference \cite{bh1} Brody and Hughston generalised this extended quantum dynamics 
by defining the stochastic Schr\"odinger evolution (hereafter the SSE)
in the case where the state space of the quantum system is given by a general
K\"ahler manifold ${\mathcal {M}}$. In this theory observables are defined by 
holomorphic Killing vectors on ${\mathcal{M}}$, that is to say vector fields 
that preserve the complex structure $J$ as well as the metric $g$. 
Such a Killing vector determines a Hamiltonian function, up to an 
additive constant. One of these Hamiltonian functions is then 
taken to be the Hamiltonian $H$ that determines the SSE. 

The SSE 
is an SDE on ${\mathcal{M}}$ with the property that its drift reduces 
the dispersion 
$V$ of $H$ and its volatility is in the direction 
of the gradient vector field $\nabla^aH$. In particular, the 
stochastic Schr\"odinger evolution is defined
in such a way that the random process for $H$ 
is a martingale. Then it turns 
out that $V$ is a supermartingale, or equivalently that the evolution 
reduces the expectation of $V$, if a certain holomorphic sectional 
curvature $K_H$ is positive. If $K_H>0$, then the expectation of $V$ is
reduced to zero asymptotically in the limit of large time, and 
Brody and Hughston \cite{bh1} call 
the corresponding evolution on the state manifold ${\mathcal {M}}$ 
a `collapse process' for the observable $H$. If $F$ is another observable, in
the sense of being the Hamiltonian for another holomorphic Killing 
vector, which has zero Poisson bracket with $H$ then the evolution  
reduces the expectation of the dispersion $V^F$ of $F$ if a certain holomorphic bisectional 
curvature is positive (these terms will be defined below). Refining
the terminology of \cite{bh1}, we shall call this a 
`reduction process' for $F$ which will be called a `collapse process' if
the expectation of 
$V^F$ is reduced to zero asymptotically in time.

The Fubini-Study metric on ${\bf CP}^n$ has positive holomorphic 
bisectional curvature. It follows that the SSE 
on ${\bf CP}^n$ is a collapse process for any choice of holomorphic
Killing vector, with Hamiltonian $H$ say, and that this is 
simultaneously a reduction process for
any other observable commuting with $H$. It is natural to
consider other K\"ahler manifolds, beginning with low dimensional or 
otherwise familiar cases, to seek further examples of state manifolds 
admitting collapse and reduction processes, and that 
is the purpose of this article. We also consider 
the L\"uders postulate, described in the next paragraph, which is 
known to be a theorem for the 
SSE in ${\bf CP}^n$ (see \cite{abbh}),
to see if it can be interpreted in these examples, and, if so, 
if it still holds. 

The L\"uders postulate in
standard quantum mechanics covers the situation when a wave function is a
linear combination of eigenstates of a Hamiltonian, at least one of
which corresponds to a degenerate eigenvalue. Suppose for simplicity
that we have a finite-dimensional state space and an initial wave-function 
$\psi(0)=a_1\psi_1+a_2\psi_2$, where $\psi_1$ lies in a one-dimensional
eigenspace, $U$ say, which therefore defines a point of the state-manifold 
${\bf CP}^n$, while $\psi_2$ lies in a two-dimensional subspace, $W$ say,which
therefore 
defines a complex projective line in the state manifold. The L\"uders
postulate (\cite{lud}) is that the measurement process 
collapses $\psi$ to $\psi_1$
with probability $|a_1|^2$ or to $\psi_2$ with probability $|a_2|^2$
despite the fact that states orthogonal to $\psi_2$ have the same
energy as it. Thus one could write $\psi_2=(\psi_3+\psi_4)/\sqrt2$
say where $\psi_3$ and $\psi_4$ were orthogonal vectors in $W$. Now
$\psi_3$ appears in the orthogonal expansion of $\psi(0)$ but there is
zero probability of collapse to it. In the SSE on 
${\bf CP}^n$, the evolution is confined 
to the projective line
through $\psi(0)$ and $\psi_1$ and so can only terminate at one of
$\psi_1$ and $\psi_2$; thus the L\"uders postulate is a theorem in
this case. If this is to be true in other state-manifolds,
then we shall need to identify some geometrical objects equivalent to these 
various structures, in particular
the submanifold to which the evolution is confined.

The plan of the article is as follows. In Section 2,  we shall review 
the relevant background material on stochastic reduction on K\"ahler 
manifolds that appeared in \cite{h1} and \cite{bh1}. In Section
3, we consider one-dimensional K\"ahler manifolds, that is to say,
manifolds of two real dimensions. Here the requirement that the SSE
determines 
a collapse process is strong enough to restrict the geometry quite 
severely: the metric is determined by a single function of one variable 
satisfying a convexity condition. In Section 4, we consider the 
two-dimensional case. Here there is room for several observables, in 
that one can have several commuting Killing vectors. A particularly 
interesting case is that of metrics with $U(2)$ symmetry transitive 
on hypersurfaces, or of LRS Bianchi-type IX in the language of general 
relativity. This case includes some familiar K\"ahler metrics, for example 
the Eguchi-Hansen metric \cite{eh} and some metrics of Hitchin 
\cite{hit}. In Section 5, we generalise some of the examples of 
Section 4 to find K\"ahler metrics with $U(N)$ symmetry admitting 
collapse processes with many observables.\\

\section{Stochastic reduction on K\"ahler manifolds}
In this section, we shall review the ideas of references 
\cite{bh1} and \cite{h1}. As we have 
said, the stochastic Schr\"odinger evolution (or SSE) defined in 
these references is an SDE for a random process with 
values in a K\"ahler manifold
${\mathcal{M}}$. The idea that the state space of quantum theory can
be generalised to a K\"ahler manifold with symmetries was introduced
by Kibble \cite{k}, and has since been developed further by a number
of authors (see, e.g., the references cited in \cite{bh1} and
\cite{h95}). Suppose that the manifold ${\mathcal{M}}$ has metric $g$, complex 
structure $J$ and K\"ahler form $\omega$. The evolution requires 
for its definition a holomorphic Killing vector $T$. The Killing
property of $T$ is
\[
{\mathcal{L}}_Tg=0
\]
and the holomorphic property is
\begin{equation}
{\mathcal{L}}_T\omega=0.
\end{equation}
\label{one}
It follows from (\ref{one}) that
\begin{equation}
i_T\omega=dH \label{two}
\end{equation}
for some function $H$, which is by definition the Hamiltonian 
function associated with 
the Killing vector $T$. In the quantum mechanical interpretation, 
$H$ is an observable and its dispersion $V$ is defined as 
\begin{equation}
V=g^{ab}\nabla_aH\nabla_bH.
\label{three}
\end{equation}
In the case of the Fubini-Study manifold it can be shown that $V$ is
the familiar squared uncertainty of $H$ in the state corresponding to
the given point of ${\mathcal {M}}$. The SSE of \cite {bh1} and \cite{h1} 
is given by
\begin{equation}
dx^a_t=(2\omega^{ab}\nabla_bH-\frac{1}{4}\sigma^2\nabla^aV)dt+\sigma\nabla^aHdW_t,
\label{SSE}
\end{equation}
where $x^a_t$ represents a random variable labelled by $t$ and taking 
values in ${\mathcal{M}}$, $dx^a_t$ is a covariant Ito differential and 
$\sigma$ is a constant. The SDE (\ref{SSE}) is defined with reference
to a fixed probability space and filtration, with respect to which
$W_t$ is a standard Brownian motion. We note in particular that
(\ref{SSE}) is covariant; in fact, the
indices are abstract \cite{h1}, and it reduces to 
the usual Schr\"odinger equation when $\sigma$ vanishes. If $\sigma$ 
does not vanish then the SDE (\ref{SSE}) has the property that it 
reduces the expected value of the dispersion 
$V$, as we shall see. The argument is as follows. First, an
application of Ito's
lemma shows that the random process $H_t=H(x_t)$ is a martingale:
\[
dH_t=\sigma V_tdW_t.
\]
where $V_t=V(x_t)$. From (\ref{SSE}), Hughston and Brody \cite{bh1}
derive an equation for $V_t$ given by
\begin{equation}
dV_t=-\sigma^2K_HV^2_tdt+\sigma\nabla^aH\nabla_aVdW_t,
\label{Veqn}
\end{equation}
where $K_H$ is a particular holomorphic sectional curvature, namely
\begin{equation}
K_H=
\frac{1}{V^2}R_{apcq}J^p_{\,b}J^q_{\,d}\nabla^aH\nabla^bH\nabla^cH\nabla^dH.
\label{KH}
\end{equation}
Now if $K_H$ is strictly
positive , then the SSE (\ref{SSE}) leads via (\ref{Veqn}) to a 
supermartingale condition on $V_t$, so that the expectation of $V_t$
decreases. 
In this case it can be shown that the evolution reduces the expectation of 
$V_t$ to zero (in the limit as $t\rightarrow\infty$) and halts at a 
critical point of $H$ or equivalently at a fixed point of
$T$. Following Brody and Hughston \cite{bh1} we shall therefore call this 
a collapse process.

Next suppose that $F$ denotes another observable. Suppose, in other words, 
that there is another holomorphic Killing vector $X$ with Hamiltonian 
function $F$. We say that $F$ and $H$ commute iff their Poisson
bracket 
vanishes, that is to say if and only if
\[
\omega^{ab}\nabla_aF\nabla_bH=0,
\]
from which it follows that the corresponding Killing vectors $X$ and 
$T$ commute (\cite{bh1}). In this case, one can ask whether a collapse process for 
$H$ necessarily also reduces or even collapses $F$. In particular,
suppose we now write 
\[V^F=g^{ab}\nabla_aF\nabla_bF\]
for the dispersion of $F$, and $V^H$ for the dispersion of $H$ given
in (\ref{three}). Then for the process $V^F$ one finds the SDE
\begin{equation}
dV^F_t=-\sigma^2K_{FH}V^F_tV^H_tdt+\sigma\nabla^aH\nabla_aV^FdW_t,
\label{VF}
\end{equation}
where $K_{FH}$ is the biholomorphic sectional curvature given by
\begin{equation}
K_{FH}=
\frac{1}{V^FV^H}R_{apcq}J^p_{\,b}J^q_{\,d}\nabla^aH\nabla^bH\nabla^cF\nabla^dF.
\label{KFH}
\end{equation}
If this biholomorphic sectional curvature is positive then the SSE,
via (\ref{VF}), necessarily reduces the expectation of the dispersion of 
$F$. We shall call this a {\em reduction} process for $F$. If we have a 
collapse process for $H$ that is simultaneously a reduction process
for $F$ but is also such that
the expectation of $V^F$ tends to zero asymptotically, we shall call
this a {\em collapse} process for the observable 
$F$. In the terminology of random
processes, we have assumed that $V^H$ is a supermartingale, and we
have deduced that it is
then necessarily a
{\em potential} (\cite{abbh}, \cite{mey}). To have simultaneously a reduction process for $F$,
$V^F$ must be a supermartingale, while to have a collapse process for $F$,
$V^F$ must also be a potential.

A process that simultaneously
collapses $H$ and $F$ must terminate at a point that is a critical
point of both $H$ and $F$. A process that collapses $H$
but merely reduces $F$ must terminate at a degenerate critical point of
$H$. It is possible, even in the case of ${\bf CP}^2$ with the
Fubini-Study metric, 
to have a process that collapses $H$ and reduces but does not
collapse $F$. To see this one can take an observable 
$H$ with a one-dimensional eigenspace,
which defines a point, say $U_1$, in ${\bf CP}^2$, and a two-dimensional
eigenspace, which comes from a degenerate eigenvalue and defines a
complex projective 
line, say $W$, in ${\bf CP}^2$. Now suppose $F$ is an observable that commutes
with $H$ but has nondegenerate eigenvalues. Recall that an eigenvector
of $F$ thought of as a $GL(3,{\bf C})$ matrix that corresponds to a 
nondegenerate eigenvalue is necessarily an eigenvector of $H$ (since
$F$ and $H$ commute by assumption). Thus the eigenspaces of $F$ will
define points $U_2$ and $U_3$ on $W$, together with $U_1$. Now the SSE
will evolve an initial state that is not on either of the complex projective
lines joining $U_1$ to $U_2$ and $U_1$ to $U_3$ either to $U_1$ or to
a point in $W$ that does not correspond to an eigenvector of $F$ and
so is not a critical point of $F$. This is a collapse of $H$ and a
reduction but not a collapse of $F$. We shall see in Case 1 of Section 4
that this phenomenon
can continue to occur in the new examples of state-manifolds explored below.

In the following sections, we shall be interested in 
K\"ahler manifolds of low dimension 
with enough holomorphic Killing vectors to give 
interesting sets of observables, and with conditions of positivity on the 
curvature sufficient to lead to collapse and reduction processes.

\section{Reduction processes on one-dimensional state manifolds}
Any Riemannian manifold of two real dimensions defines a 
K\"ahler manifold of one complex dimension. Here there is only one 
independent component of curvature, namely the Gauss curvature. To
have a collapse process this must be positive. It is a theorem of 
Cohn-Vossen (for references see \cite{cv}) that a complete Riemann 
surface with positive Gauss curvature is diffeomorphic to a sphere or 
a plane. If we assume, as seems reasonable, that a state manifold must 
be complete, then the state manifold for a collapse process is 
necessarily one of these two.

To define the SSE we need a holomorphic 
Killing vector, which in the case of a one-dimensional state manifold 
is any Killing vector. This could 
have open or closed trajectories on the plane but on the sphere must 
have closed trajectories. In either case the metric can be written in the 
form
\begin{equation}
ds^2=d\theta^2+(S(\theta))^2d\phi^2
\label{met2}
\end{equation}
in terms of a function $S(\theta)$, where the Killing vector 
is $T=\partial/\partial\phi$. For this metric 
we calculate the Gauss curvature and obtain
\begin{equation}
K= -\frac{1}{S}\frac{d^2S}{d\theta^2}.
\label{K}
\end{equation}
For a collapse process this needs to be 
positive, and it then follows that $S$ must have a zero. For
completeness of the state manifold at a zero of $S$ the trajectories 
must be closed which makes $\phi$ periodic, and for definiteness we 
assume the period to be $2\pi$. Also for completeness we need to avoid 
having a conical singularity at any zero of $S$, which requires
$dS/d\theta=\pm1$ there. This is because, for an infinitesimal circle around 
the zero, we want the ratio of circumference to radius in the limit of 
vanishing radius to be $2\pi$. For a metric on the plane, $S$ has a 
single zero; while for a metric on the sphere, $S$ has two zeroes. For 
definiteness, we suppose that the necessary zero of $S$ is at $\theta=0$.

The K\"ahler form associated with the metric (\ref{met2}) is given by
\[
\omega=S(\theta)\,d\theta\wedge\phi.
\]
Then as a consequence of (\ref{two}) the Hamiltonian $H(\theta)$ satisfies
\begin{equation}
\frac{dH}{d\theta}=-S.
\label{H}
\end{equation}
It then follows from (\ref{three}) that the dispersion $V$, is given by
\begin{equation}
V=S^2.
\label{V}
\end{equation}
The zero or zeroes of $S$ are fixed points of the Killing vector $T$ 
and critical points of the Hamiltonian $H$. The SSE will evolve towards such a zero.

Looking ahead to Section 5, we introduce a 
holomorphic coordinate tied to the symmetry and taking the form 
$z=re^{i\phi}$. The K\"ahler metric (\ref{met2}) necessarily has 
a K\"ahler potential $\Sigma(u)$ where $u=r^2$. In terms of $\Sigma$ 
the metric can be written
\[
ds^2=2\Sigma_{,z\bar{z}}dzd\bar{z},\label{met3}\]
and thus
\[
    ds^2=2(\dot{\Sigma}+u\ddot{\Sigma})(dr^2+r^2d\phi^2),\]
where the dot denotes $d/du$. Comparing (\ref{met3}) with (\ref{met2})
and (\ref{V}) we 
can make the identification
\begin{equation}
V=2u(\dot{\Sigma}+u\ddot{\Sigma}). 
\label{V2}
\end{equation}
Then as a consequence of (\ref{H}) and (\ref{V}) we deduce that 
$H=-u\dot{\Sigma}$ and $V=-2u\dot{H}$.

The zero of $S$ at $\theta=0$ we can suppose to correspond to a zero 
of $V$ at $r=0$. It is convenient to introduce a new radial coordinate 
$\chi=\log r$ so that the metric takes the form
\[
ds^2=V(d\chi^2+d\phi^2). 
\]
With these coordinates, the SSE (\ref{SSE}) becomes
\begin{equation}
d\chi_t = -\frac{1}{2}\sigma^2\frac{d\log V}{d\chi}dt+\sigma
dW_t,\;\;\;d\phi = 2dt.
\label{SSE2} \\ 
\end{equation}
Note that the evolution for $\phi$ is deterministic: this will
be important later, in Section 5. In terms of the holomorphic coordinate $z$, we find
\begin{equation}
dz=z[2i+\frac{1}{2}\sigma^2(1-\frac{d\log V}{d\chi})]dt+z\sigma dW_t.
\label{zeqn}
\end{equation}
This is an equation that will be generalised in Section 5. Finally the 
Gauss curvature (\ref{K}) is given by
\begin{equation}
K=-\frac{1}{2V}\frac{d^2 \log V}{d\chi^2}.
\label{K2}
\end{equation}
Comparing (\ref{SSE2}) and (\ref{K2}) we see that if the curvature 
$K$ is positive
then the drift in the 
SSE for $\chi$ is monotonic and increasing 
in $\chi$. 

By assumption there is a fixed point at $r=0$ or $\chi=-\infty$. Here 
$S=0$ and $dS/d\theta=1$, and one can calculate the limit of the drift in
$\chi$ in (\ref{SSE2}) as $-\sigma^2$. The drift is 
therefore towards the fixed point. 
Now consider increasing $\chi$. If $dS/d\theta$ ever has a zero then the 
positivity of $K$ forces $S$ to have another zero and completeness 
then requires $dS/d\theta=-1$ at this zero. It easy to see that
this zero occurs at $\chi=\infty$. The manifold is now necessarily 
the sphere, and the limit of the drift near this fixed point is 
$\sigma^2$, so again it is towards the fixed point. The drift 
changes sign at the maximum of $V$. If $dS/d\theta$ never has a zero then 
the drift never changes sign and for all $\chi$ is towards the fixed 
point at $\chi=-\infty$.

By calculating specific examples of surfaces of revolution in
Euclidean three-space one finds that for large positive 
$\chi$ on an asymptotically 
hyperbolic surface the drift tends to $-k\sigma^2$ where $k$ is a 
positive constant less than unity, while for an asymptotically parabolic 
surface the drift tends to zero as $O(\chi^{-1})$.

In the case when the manifold is a sphere, it is possible to calculate
the probabilities $\pi_+$ and $\pi_-$, that the evolution terminates at
$\chi=\infty$ or at $\chi=-\infty$ respectively, in terms of the 
initial value of $\chi$. This is done by solving the backward
Fokker-Planck equation associated with the SDE (\ref{SSE2})
with the relevant boundary conditions (\cite{book}). The result for
$\pi_+$ is
\[\pi_+=\frac{H(\chi)-H(-\infty)}{H(\infty)-H(-\infty)}\]
with $\pi_-=1-\pi_+$. 

For the asymptotically hyperbolic or parabolic
(or even cylindrical) examples, the same calculation gives $\pi_-=1$,
as expected . In other words, the reduction proceeds towards the
single `eigenstate', ultimately resulting in collapse.

\section{Reduction processes on two-dimensional state manifolds}

In two dimensions there are many more curvature components, and as a
consequence the 
restriction that just one holomorphic sectional curvature should be positive 
is not a strong one. Simple examples of K\"ahler manifolds 
with this property can be constructed by 
taking products of a Riemann surface with an example from the previous 
section. More interesting examples arise if we assume that there are
more observables and look for processes that collapse or reduce several of
these observables at once. With this possibility in view, we consider metrics with a $U(2)$ 
action transitive on three-surfaces, equivalently stated as an action with
three-dimensional principal orbits. That is to say, we have holomorphic 
Killing vectors $T$ and $X_i$, for $i=1,2,3,$ with the commutators
\begin{equation}
[X_i,X_j]  = -\epsilon_{ij}^{\;\;\;\;k} X_k;  \;\;\; [X_i,T]=0.
\end{equation}
We shall thus be led to examine explicit examples of nonlinear state
manifolds with observables $H$ and $S_i$ 
associated with the symmetries $T$ and $X_i$ respectively, which we 
can think of as {\emph{energy}} and {\emph{spin}} respectively. 
We shall consider the 
evolution defined by $T$ and seek manifolds that give rise to collapse 
processes for the energy. It will turn out that some of these 
simultaneously reduce the spin.

We begin by introducing a set of Euler angles $(\theta,\phi,\psi)$, in terms of
 which we shall take the Killing vectors to be given by the 
following expressions:
\begin{eqnarray}
X_1+iX_2 & = & e^{i\phi}(i\frac{\partial}{\partial\theta} 
-\cot\theta \frac{\partial}{\partial\phi}
+\csc\theta\frac{\partial}{\partial\psi}), \label{KVs} \\
X_3 & = & \frac{\partial}{\partial\phi},\nonumber \\
T & = & \frac{\partial}{\partial\psi}. \nonumber
\end{eqnarray}
We shall work with the usual basis of invariant one-forms $\sigma_i$, which
are given in 
these coordinates by
\begin{eqnarray}
\sigma_1+i\sigma_2 & = & e^{i\psi}(d\theta-i\sin\theta d\phi),\label{forms}\\ 
\sigma_3 & = & d\psi+\cos \theta d\psi, \nonumber
\end{eqnarray}
so that $d\sigma_1=\sigma_2\wedge\sigma_3$,
 $d\sigma_2=\sigma_3\wedge\sigma_1$ and 
 $d\sigma_3=\sigma_1\wedge\sigma_2$. The most general metric with the desired symmetry is now
\begin{equation}
ds^2=dt^2+a^2(t)(\,\sigma_1^{\,2}+\sigma_2^{\,2})+c^2(t)\sigma_3^{\,2}.
\label{met5}
\end{equation}
The metric depends on a pair of 
functions $a(t)$ and $c(t)$, where $t$, which we can 
think of as either a `time' or a `radial' coordinate, labels the surfaces of 
homogeneity. The generic surface of homogeneity, or equivalently the 
principal orbit of the symmetry group $U(2)$, can be a three-sphere or a
Lens space $L(3,n)$ which is a quotient of the three-sphere by a cyclic
group of order $n$, and the 
orbits degenerate at zeroes of $a$ and/or $c$. Given the topology of
the principal orbit, the topology of the
underlying manifold ${\mathcal{M}}$ is determined by the 
nature of the degenerate
orbits and the (one-dimensional) manifold say ${\mathcal{T}}$ which $t$ ranges
over. Clearly this must be one of four possibilities: a circle, an
interval, the whole line or a half-line. The first two give compact
state-manifolds and the last two give noncompact state-manifolds. 

It will be convenient 
to work with a particular orthonormal frame for this metric, namely
\begin{equation}
\theta^0=dt,\;\theta^1=a\sigma_1,\;\theta^2=a\sigma_2,\;\theta^3=c\sigma_3.
\label{frame}
\end{equation}
In terms of this frame we define the complex structure $J$ by
\[
J\theta^0=\theta^3,\;J\theta^1=\theta^2.
\]
The corresponding two-form is then given by
\begin{equation}
\omega=cdt\wedge\sigma_3+a^2\sigma_1\wedge\sigma_2.
\label{Kah}
\end{equation}
It can be checked that the complex structure is 
automatically integrable and preserved by the 
Killing vectors, so that these are all holomorphic. The K\"ahler
condition is that the form $\omega$ given by (\ref{Kah}) should be
closed. This requires that
\begin{equation}
\frac{d(a^2)}{dt}=c.
\label{a}
\end{equation}
Equation (\ref{a}) immediately shows that ${\mathcal{T}}$ cannot
be a circle: if $a$ were periodic in $t$ then $c$ would have zeroes
at which the metric would degenerate. On the other hand, since
$c^2=g(T,T)$, we need at least one zero in $c$ or there are no fixed
points of the evolution. This shows that ${\mathcal{T}}$ must be
either an
interval, in which case there will be two critical values of $H$, or a
half-line, when there will be just one.

We next find Hamiltonian functions for the four Killing vectors. Let
us denote these 
Hamiltonian functions as $H$ for $T$ and $S_i$ for $X_i$. Then from
(\ref{KVs}), (\ref{forms}) 
and (\ref{Kah}) these are easily found as
\begin{eqnarray*}
H & = & -a^2, \nonumber \\
S_1+iS_2 & = & -a^2\sin \theta e^{i\phi}, \label{Hams}\\
S_3 & = & -a^2\cos \theta. 
\end{eqnarray*}
As a consequence, we note that
\[
H^2=S_1^{\,2}+S_2^{\,2} +S_3^{\,2}.
\]
With $\omega$ as in (\ref{Kah}) it is a simple matter to check that
the Poisson brackets $\omega(T,X_i)$ all vanish, so that the
observables $S_i$ commute with the Hamiltonian $H$. The $S_i$ have the 
usual $SU(2)$ commutators with each other, i.e. $\omega(S_1,S_2)=S_3$, 
$\omega(S_2,S_3)=S_1$ and $\omega(S_3,S_1)=S_2$.

It is convenient to introduce 
a new radial coordinate $R=2a$, which is allowed because
$\dot{a}\neq0$ in the interior of the coordinate range (otherwise $c$
would have a zero, which can only happen at the end of the range). 
Then $c=\frac{1}{2} R\dot{R}$ and the metric 
(\ref{met5}) can be written in the form
\begin{equation}
ds^2 = \frac{1}{F}dR^2+\frac{1}{4}R^2
(\sigma_1^{\,2}+\sigma_2^{\,2})+\frac{1}{4}R^2F\sigma_3^{\,2},
\label{met6}
\end{equation}
where $F(R)=\dot{R}^2$. Written like this, the metric depends on one 
function $F$ of $R$. There are various possibilities for the topology 
of the underlying manifold, determined by the range of $R$ for which
$F$ is positive and the behaviour of $F$ near its zeroes. There is 
some standard terminology too: see, for example, reference \cite{mjp}. If $R=0$ is in 
the allowed range then for completeness of the metric we need $F=1$ 
there: this is the condition that the coordinate singularity, which 
is like the singularity at the origin of polar coordinates, can be 
removed and the manifold can be completed by inserting a point. A 
coordinate singularity of this kind is called a `nut' in the
literature. 

For completeness at a zero of $F$, say at $R=R_0\neq 0$, we need 
\begin{equation}
R_0F'(R_0)=\pm2n
\label{bolt}
\end{equation}
for a positive integer $n$, where the principal orbit is $L(3,n)$, so
$n=1$ if the principal orbit is a three-sphere. In this case the 
coordinate singularity 
can be removed and the manifold completed by inserting a two-sphere. A 
coordinate singularity of this kind is called a `bolt' in the
literature. If there are two bolts then for completeness 
they must have the same $n$
with one plus and one minus in (\ref{bolt}). 
If there is a nut and a bolt then the principal orbits are three-spheres
and $n=1$ at the bolt. There cannot be two nuts because between them
$\dot{a}$ would have a zero whence so would $c$ by (\ref{a}) and there
would be a bolt before the second nut. We shall see 
below how the possibilities are constrained further by conditions of
positivity on the curvature.\\

We can calculate the dispersions $V$ and $V_i$ associated with the 
observables $H$ and $S_i$ respectively by use of (\ref{KVs}) and 
(\ref{met6}). These turn out to be
\begin{eqnarray}
V & = & \frac{1}{4}R^2F, \nonumber\\
V_1 & = & \frac{1}{4}R^2(\sin^2\theta + F\cos^2\theta),\nonumber\\
V_2 & = & \frac{1}{4}R^2(\cos^2\phi + \sin^2\phi(\cos^2\theta + 
F\sin^2\theta)), \nonumber\\
V_3 & = & \frac{1}{4}R^2(\sin^2\phi + \cos^2\phi(\cos^2\theta + 
F\sin^2\theta)). \label{norms}
\end{eqnarray}
The fixed points of the Killing vectors, which are the zeroes of the 
respective dispersions, can be read off from (\ref{norms}). All four 
Killing vectors vanish at the nut, if there is one. Thus a nut is an
isolated fixed point of $T$, and therefore a non-degenerate critical point of
$H$, as well as an isolated fixed point of each $X_i$. The observables
$H$ and $S_i$ all vanish there. If there is a
bolt, 
then $T$ vanishes at all points of it while each $X_i$ vanishes at a 
pair of antipodal points, and for varying~$i$ the pairs are 
symmetrically arranged. Thus a bolt is a whole two-sphere, in fact a
${\bf CP}^1$, of fixed
points of $T$ that are degenerate critical points of $H$, and each
$S_i$ has two critical values on the bolt. The Hamiltonian takes the
value $H_0=-R_0^2/4$ on a bolt at $R=R_0$ and the critical values of
the $S_i$ are $\pm H_0$.\\

We next need to calculate the Riemann tensor for the metric 
(\ref{met6}). This is readily done in the orthonormal basis of 
(\ref{frame}), with the following result:
\[R_{0101}  =  R_{0202} =  -\frac{1}{2R}F',\]
\[R_{0123}  =  R_{0231} =  -\frac{1}{2}R_{0312} = \frac{1}{2R}F',\]
\[R_{0303}  =  -\frac{1}{2}(F''+\frac{3}{R}F'),\]
\begin{equation}
R_{1212}  =  \frac{4}{R^2}(1-F),
\label{Riem}
\end{equation}
where the prime denotes $d/dR$. Once we have the Riemann tensor, 
we can calculate the biholomorphic 
sectional curvatures. Suppose that $U$ and $W$ are arbitrary unit 
vectors, so that
\begin{eqnarray*}
U & = & A e_0+Be_1+Ce_2 +De_3,\\
W & = & \alpha e_0+\beta e_1+\gamma e_2 +\delta e_3, 
\end{eqnarray*}
where $A^2+B^2+C^2+D^2=\alpha^2 +\beta^2+\gamma^2+\delta^2=1$ and 
the $e_i$ are the basis of vector fields dual to the $\theta^i$. Then
\begin{eqnarray*}
JU & = & -D e_0-C e_1+B e_2 +A e_3,\\
JW & = & -\delta e_0-\gamma e_1+\beta e_2 +\alpha e_3,
\end{eqnarray*}
and the corresponding biholomorphic sectional curvature can be written
\begin{eqnarray}
R(U,JU,W,JW)&=&R_{0303}(\alpha^2+\delta^2)(A^2+D^2)\nonumber\\
&&+R_{0312}((\alpha B+\beta A-\gamma D-\delta C)^2\nonumber\\
&&        +(\alpha C+\beta D+\gamma A+\delta B)^2)\nonumber\\
&&+R_{1212}(\beta^2+\gamma^2)(B^2+C^2).
\label{holsec}
\end{eqnarray}
For the Killing vector $T=ce_0$ that determines the evolution, the 
relevant holomorphic sectional curvature $K_H$ from (\ref{KH}) and 
above is $R_{0303}$. The condition for a collapse process 
for $H$ is therefore
\begin{equation}
R_{0303}=-\frac{1}{2R^3}(R^3F')'>0.
\label{pos1}
\end{equation}
If we want the condition for this process to 
reduce, say, $S_3$ as well, we first need $X_3$ 
in the basis $(e_i)$. This is
\[
X_3 = a\sin \theta(\sin\psi e_1-\cos \psi e_2)+c\cos\theta e_3.
\]
Now for the biholomorphic sectional curvature, as defined by 
analogy with (\ref{KFH}) we find
\[
K_{HS_3} = \frac{1}{\Delta}(R_{0303}F\cos^2\theta+R_{0312}\sin^2\theta),
\label{pos2}
\]
where $\Delta=F\cos^2\theta+\sin^2\theta$.

We shall have a reduction process for $S_3$ as well as for $H$ if this is 
positive too, which requires
\begin{equation}
R_{0312}=-\frac{1}{2R}F'>0
\label{pos3}
\end{equation}
in addition to (\ref{pos1}). We would obtain the same condition by 
considering $S_1$ or $S_2$ instead of $S_3$, so that 
(\ref{pos1}) and (\ref{pos3}) 
taken together are necessary and sufficient for the evolution defined 
by $T$ to collapse $H$ and simultaneously reduce any one of the $S_i$.

We obtain a range of examples by considering various cases, which we
now proceed to summarise.

\subsection*{Case 1: one nut and one bolt}
Suppose that $F$ is positive at $R=0$, has a zero at $R=R_0>0$, and is
positive in between, so that the 
state manifold ${\mathcal{M}}$ is compact. For completeness we need a 
nut at $R=0$, so that $F(0)=1$, $F'(0)=0$, and the principal orbit is a
three-sphere, and a bolt at $R=R_0$, 
so that $R_0F'(R_0)=-2$. We can now identify ${\mathcal{M}}$ as 
topologically (and therefore in 
fact biholomorphically) equivalent to ${\bf CP}^2$.

For a collapse process for $H$ we need (\ref{pos1}), 
but since $F'(0)=0$ this will 
ensure that $F'(R)<0$, from which (\ref{pos3}) will necessarily
follow: by insisting on a collapse process for $H$ we automatically obtain 
a reduction process for $S_3$. Finally, $F$ is decreasing with $F(0)=1$ so 
that $F<1$ and by (\ref{Riem}) this ensures that $R_{1212}>0$. 
Putting this together with (\ref{holsec}) we now have enough to make 
every biholomorphic sectional curvature positive.
It is a theorem of Siu and Yau (\cite{s1}, see also \cite{m}) that a
 compact K\"ahler manifold with positive biholomorphic sectional 
curvatures is necessarily ${\bf CP}^n$.

The reduction process for $S_3$ will not necessarily be a collapse
process as we shall see from a discussion of the L\"uders postulate in
Section 5. However, since every biholomorphic sectional
curvature is positive given
(\ref{pos1}), we could instead use $S_3$ as Hamiltonian. This new SSE would
give a collapse process for $S_3$ which would also necessarily reduce
$H$. Now the critical points of $S_3$ are nondegenerate and therefore are
also critical points of $H$, as we can see in any case from
(\ref{norms}). Thus the reduction process for $H$ is in fact a
collapse process for $H$ and this process simultaneously collapses
$S_3$ and $H$.

This example, with a metric on ${\bf CP}^2$ that is different from
the Fubini-Study metric, is very similar to the
example with a degenerate $H$ discussed in the Section 1, and can be
seen as a natural generalisation of this example.

\subsection*{Case 2: two bolts}
The other compact case arises if $F$ has zeroes at two nonvanishing values
 of $R$ and is positive in between. Call these $R_0$ and $R_1$ 
with $R_0\leq R_1$. Then we have 
two bolts and we require that
\begin{equation}
R_0F'(R_0)=-R_1F'(R_1)=2n
\label{bolts}
\end{equation}
for some positive integer $n$, so that the principal orbit is
$L(3,n)$. Looking at (\ref{Riem}) we see that 
$R_{0312}$ must change sign, so that by (\ref{pos3}) we cannot have 
a reduction process for $S_3$ even though, by (\ref{pos1}), we can 
arrange to have a collapse process for $H$. Note also that the
critical points of $H$ lie on two distinct complex projective lines,
the two bolts. This could not happen in $\bf{CP}^2$ where two complex
projective lines necessarily meet.

There are some specific examples of this case due to Hitchin 
\cite{hit}. In our notation they are defined by
\[
F(R)=\frac{(R^2-1)(sn+1-R^2)}{sR^2},
\]
where $s$ is a positive real constant and $n$ is a positive integer 
greater than $1$. There are bolts at the two  
positive zeroes of $F$ and (\ref{bolts}) is satisfied with the same 
$n$ at both. The state manifold ${\mathcal{M}}$ is a rational surface, 
specifically the projective space ${\bf{P}}({\mathcal{O}}(n)+
{\mathcal{O}}(0))$ of the vector bundle ${\mathcal{O}}(n)+
{\mathcal{O}}(0)$ over ${\bf CP}^1$. 

From (\ref{pos1}) we find $R_{0303}=4/s$ which is positive, 
so that we do have a 
collapse process for $H$, but as was already noted, this process does 
not simultaneously reduce any of the $S_i$. The interest of this 
example for Hitchin was that for small enough $s$, in fact for 
$s<1/n^2$, the holomorphic sectional curvatures are all positive 
(although the biholomorphic sectional curvatures are not, as can 
be seen from the theorem of Siu and Yau \cite{s1} already cited).\\

Putting these two cases together, we can assert that for a compact 
${\mathcal{M}}$ with the symmetry assumed here, 
if the stochastic Schr\"odinger evolution collapses 
$H$ and reduces $S_3$ then the manifold must be ${\bf CP}^2$ with a metric of
positive biholomorphic sectional curvature.

\subsection*{Case 3: semi-infinite with a nut}
The simplest noncompact case has a nut at $R=0$, for which therefore $F(0)~=~1, 
F'(0)~=~0$, and $F$ is everywhere positive. If we have a collapse process 
for $H$ then, by (\ref{pos1}), we must have $(R^3F')'~>~0$. With the boundary
conditions from the nut this forces $F'<0$ and so $0<F<1$ for all
positive $R$. The state manifold in this case is ${\bf C}^2$.

Now from (\ref{Riem}) and (\ref{holsec}) all biholomorphic sectional 
curvatures are positive. In particular the collapse process for $H$ 
is simultaneously a reduction process for all the $S_i$. Further, this
reduction process is also a collapse process because the nut is a
nondegenerate critical point for all the observables. However this case is 
less interesting because the endpoint of the stochastic evolution is 
the origin, where all the observables are zero.

A simple example of this case is given by
\[
F=\frac{1+\lambda R^2}{1+R^2},
\]
where $\lambda$ is a real constant with $0\leq \lambda \leq 1$. The 
corresponding metric is the flat metric on ${\bf C}^2$ for $\lambda=1$, 
is asymptotic to a deformed (`Berger') three-sphere at infinity if 
$0<\lambda<1$ and is ALF (or asymptotically locally flat) in the 
language of \cite{mjp} if $\lambda=0$.

\subsection*{Case 4: semi-infinite with a bolt}
The final case has a bolt at say $R=R_0$ so that $F(R_0)=0, 
R_0F'(R_0)=2n$, with $F>0$ for $R>R_0$. We can arrange to satisfy 
(\ref{pos1}) so that we have a collapse process for $H$, but we cannot 
satisfy (\ref{pos3}) and simultaneously obtain a reduction process for 
$S_i$. A familiar example of this case is the Eguchi-Hanson metric 
(\cite{eh}, \cite{mjp}) which has $F=1-a^4/R^4$. This has a bolt with
$n=2$ at 
$R=a$ and is asymptotically locally Euclidean in the standard
terminology (see, for example, \cite{mjp}).\\

In summary, we have constructed a variety of two-dimensional quantum 
state manifolds admitting collapse
processes for energy, and some of these simultaneously reduce the 
spin. The most interesting case is the first, which corresponds to a
 metric on ${\bf CP}^2$ with positive biholomorphic sectional
curvatures but which is different from the Fubini-Study metric. In the 
next section, we consider a generalisation to metrics with $N$ complex
dimensions and  
$U(N)$-symmetry.

\section{K\"ahler metrics with U(N) symmetry}
We want to generalise the calculations of the last section to 
the consideration of K\"ahler metrics with $U(N)\,$-symmetry transitive on
hypersurfaces. To do this we 
mimic the calculation leading to (\ref{met3}). Suppose we have 
complex coordinates $z^a,a=1,\ldots,N$, where these are coordinate
indices (not abstract indices), and a K\"ahler potential 
$\Sigma(u)$ where
\begin{equation}
u=r^2=\delta_{a\bar{b}}z^a\bar{z^b}=\sum_{a=1}^{N}|z^a|^2.
\label{u}
\end{equation}
The metric in these coordinates is $ds^2=2g_{a\bar{b}}dz^ad\bar{z^b}$ with
$g_{a\bar{b}}  =  \partial^2\Sigma/
\partial z^a \partial \bar{z^b}$ so that
\begin{equation}
g_{a\bar{b}} =  \dot{\Sigma} \delta_{a \bar{b}} + 
\ddot{\Sigma}z_{\bar{a}}\bar{z}_b, \label{met7}
\end{equation}
where as before the dot means $d/du$, and we have introduced the notation
\[
z_{\bar{a}}=\delta_{b\bar{a}}z^b,\,\,\bar{z}_b=\delta_{b\bar{a}}\bar{z^a}.
\]
The metric (\ref{met7}) is positive definite provided the following
conditions hold:
\[
\dot{\Sigma}>0,\,\,\dot{\Sigma}+u\ddot{\Sigma}>0.
\label{pos4}
\]
The metric degenerates where either of these quantities vanishes; but,
as we shall see, these degeneracies may correspond to removable 
coordinate singularities.
The inverse metric is easily found to be given by
\begin{equation}
g^{a\bar{b}}=\frac{1}{\dot{\Sigma}}(\delta^{a\bar{b}}-Qz^a\bar{z^b}),
\label{met8}
\end{equation}
where $Q=\ddot{\Sigma}/(\dot{\Sigma}+u\ddot{\Sigma})$.
In these coordinates we can find the following holomorphic Killing vectors:
\begin{eqnarray}
T & = & i(z^a\frac{\partial}{\partial z^a}-
\bar{z^b}\frac{\partial}{\partial\bar{z^b}}),\\
X & = & i(z^aH_a^{\,c}\frac{\partial}{\partial z^c}-
\bar{z^b}H_{\bar{b}}^{\,\bar{d}}\frac{\partial}{\partial\bar{z^d}}),
\nonumber
\label{KV2}
\end{eqnarray}
where $H_a^{\,b}=\delta^{b\bar{a}}H_{a\bar{a}}$ and 
$H_{\bar{b}}^{\,\bar{d}}=\delta^{b\bar{d}}H_{b\bar{b}}$ for 
an arbitrary trace-free Hermitian matrix $H_{a\bar{a}}$. It is 
easy to see that these are holomorphic and they are Killing vectors since 
they preserve the function $u$ of (\ref{u}). The Killing vectors of the 
form $X$ generate the Lie algebra of $SU(N)$ as $H_{a\bar{a}}$ 
runs over trace-free Hermitian matrices. The Hamiltonians for these 
Killing vectors turn out to be 
\begin{equation}
H  =  -u\dot{\Sigma},\label{H2}
\end{equation}
and
\begin{equation}
S  =  -H_{a\bar{b}}z^a\bar{z^b}\dot{\Sigma}.
\end{equation}
We note that, interestingly, (\ref{H2}) is formally just as in the one-dimensional case. 
With the aid of (\ref{met8}) the dispersion $V$ of $H$ is found to be
\[
V=2g^{a\bar{b}}\partial_aH\partial_{\bar{b}}H=-2u\dot{H},
\]
which is also formally the same as in the one-dimensional case
(\ref{V2}). The critical points of $H$ are at the zeroes of $V$, that
is to say at $u=0$ or at $\dot{\Sigma} + u\ddot{\Sigma} =0$, and the
second of these is where the metric degenerates. 

The dispersion $V^S$ of $S$ can be written in either of the following forms:
\begin{equation}
V^S =  2g_{a\bar{b}}H_{\bar{c}}^{\,\bar{b}}\bar{z^c}H_d^az^d
\label{VS1}
\end{equation}
or
\begin{equation}
V^S =  2(\dot{\Sigma}H_{\bar{c}}^{\,\bar{b}}H_{\bar{b}d}\bar{z^c}z^d
+\ddot{\Sigma}(H_{a\bar{b}}z^a\bar{z^b})^2).\label{VS2}
\end{equation}
From (\ref{VS1}) one can see that the fixed points of $X$, which are
the critical points of $S$ and the zeroes of $V^S$, necessarily occur
either at $u=0$, where every $z^a$ is zero, or on the surface where the metric
degenerates. The latter critical points, by (\ref{VS2}), occur for
$z^a$ satisfying
\[
(\delta_{a\bar{b}}z^a\bar{z^b})
(H_{\bar{c}}^{\,\bar{e}}H_{\bar{e}d}z^d\bar{z^c})=
(H_{a\bar{b}}z^a\bar{z^b})^2,
\]
which holds when $z^a$ is (proportional to) an eigenvector of
$H_a^{\, b}$. We can assume that the eigenvalues of $H_a^{\, b}$ are
distinct (since we can choose a basis of trace-free Hermitian matrices
all elements of which have distinct eigenvalues) and then the critical
points of $S$ will be nondegenerate.  \\

We want to write down the SSE (\ref{SSE}) in these 
coordinates. For this we note with the aid of 
(\ref{met8}) that
\[
\nabla^aH   =  z^a
\]
and
\[
\nabla^a V  =  \frac{-2u\dot{V}}{V}z^a.
\]
Now (\ref{SSE}) becomes just
\begin{equation}
dz^a=z^a[2i+\frac{\sigma^2}{2}(1-\frac{2u\dot{V}}{V})]dt+z^a\sigma dW_t,
\label{SSE3}
\end{equation}
which should be compared with equation (\ref{zeqn}). Recall that the indices in
(\ref{SSE3}) are coordinate indices. To give something like (\ref{met3}),
we introduce a form of polar coordinates
\begin{equation}
 z^a=r\zeta^a \label{zeta}
\end{equation}
where $\delta_{a\bar{b}}\zeta^a\bar{\zeta^b}=1$. The metric then becomes
\begin{eqnarray}
ds^2 & = & 2g_{a\bar{b}}dz^ad\bar{z^b}\label{met9}\\
& = & 2(\dot{\Sigma}+u\ddot{\Sigma})dr^2
+2u\dot{\Sigma}\delta_{a\bar{b}}d\zeta^ad\bar{\zeta^b}
+\frac{1}{4}u^2\ddot{\Sigma}\,\Theta^2, \nonumber
\end{eqnarray}
where $\Theta$ is defined by
\begin{equation}
\Theta  =  2i\delta_{a\bar{b}}\zeta^ad\bar{\zeta^b}.
\label{theta}
\end{equation}
Equation (\ref{met9}) is the counterpart of (\ref{met3}). If we use the polar
decomposition (\ref{zeta}) in (\ref{SSE3}) we find
\begin{equation}
d\zeta^a=2i\zeta^adt, \label{zev}
\end{equation}
which is again deterministic, just as the equation for $\phi$ in
(\ref{SSE2}) was. This is the significant result for the L\"uders postulate
discussed in the Introduction. Suppose the SSE starts from an initial
point with coordinates $z^a_0$. By a linear change of coordinates,
which is a symmetry, we can arrange that $z^a_0=0$ for $a=2,\ldots,N,$
and then the evolution (\ref{zev}) will ensure that the state remains
on this one-dimensional complex submanifold. 

The other part of (\ref{SSE3}) is conveniently written
in terms of $\chi=\frac{1}{2}\log u$ and 
then another application of Ito's lemma leads to
\begin{equation}
d\chi_t=-\frac{1}{2}\sigma^2\frac{d\log V}{d\chi}dt +\sigma dW_t,
\label{SSE4}
\end{equation}
which is precisely the same as in the one-dimensional case (\ref{SSE2}).

To make the metric (\ref{met9}) more like (\ref{met6}) in appearance, we
introduce the radial coordinate $R$ by $R^2=2u\dot{\Sigma}$ to find
\begin{equation}
ds^2=\frac{1}{F}dR^2+\frac{1}{4}R^2g_{FS}+\frac{1}{4}R^2F\Theta^2, 
\label{met10}
\end{equation}
where
\[
F  =  \frac{\dot{\Sigma}+u\ddot{\Sigma}}{\dot{\Sigma}},\;\;\; g_{FS} =  
\delta_{a\bar{b}}d\zeta^ad\bar{\zeta^b}-\frac{1}{4}\Theta^2.
\]
Here $g_{FS}$ is the standard Fubini-Study metric on
${\bf{CP}}^{N-1}$, with constant holomorphic sectional curvature equal
to one, and $\Theta$ from (\ref{theta}) is real because of the normalisation of
$\zeta^a$. By transforming the metric to look like (\ref{met6}),
we make it easier to discuss the underlying topology. As before, the
range of the `time'-coordinate is an interval for a compact
state-manifold or a half-line for a noncompact state-manifold.

The next step is to find the connection and curvature, and for this 
we may use formulae from Kodaira and Morrow 
\cite{km}:
\[
\Gamma^a_{\,bc} =  g^{a\bar{a}}\partial_bg_{c\bar{a}},\;\;\;
R^a_{\;b\bar{c}d} =  \partial_{\bar{c}}\Gamma^a_{\,bd}.\]
These can be found explicitly in terms of $\Sigma$. For our 
purposes we need $K_H$ as in 
(\ref{KH}) with the Killing vector $T$ of (\ref{KV2}). This turns out to be
\begin{equation}
K_H=-\frac{1}{2V}\frac{d^2 \log V}{d\chi^2}, \label{K4}
\end{equation}
which is the same expression as (\ref{K2}). The positivity of $\dot{H}$ is 
needed for the metric (\ref{met6}) to
be positive definite so again the positivity of $K_H$ forces the drift in 
(\ref{SSE4}) to be monotonic. If we use the metric form (\ref{met10})
then the conditions of positivity turn out to be precisely what they
were in Section 4. Thus positivity of $K_H$ is again (\ref{pos1}) and
for positivity of the holomorphic bisectional curvatures $K_{HS}$ we
need as well (\ref{pos3}). Finally, all holomorphic bisectional
curvatures are positive if as well $R_{1212}$ as given in (\ref{Riem})
is positive.

Following through the analysis of nuts and bolts, we have the four
cases as before:\\ 

\noindent {\bf Case 1: one nut and one bolt} is 
a metric on ${\bf CP}^N$ with positive holomorphic
bisectional curvature, so that the SSE gives a collapse
process for the Hamiltonian that is simultaneously a reduction 
process for any one of the $SU(N)$ observables. The 
Hamiltonian has two critical
values: one (nondegenerate) occurs at a single point (the nut),
and one (degenerate) on a ${\bf CP}^{N-1}$ (the bolt). Any one of the
$SU(N)$ variables has $N$ nondegenerate critical points on the bolt
and one at the nut. The SSE confines
the evolution to a 
submanifold that is actually a complex projective line through the
nut and the initial state. Thus the L\"uders postulate holds
here. This analysis includes the case $N=2$ which we did not do
separately in Section 4. \\

\noindent {\bf Case 2: two bolts} is a metric on 
the projective space 
${\bf{P}}({\mathcal{O}}(n)+{\mathcal{O}}(0))$ of the vector 
bundle ${\mathcal{O}}(n)+{\mathcal{O}}(0)$ over ${\bf CP}^{N-1}$. The
Hamiltonian has two degenerate critical values, at the two bolts
respectively, and the L\"uders postulate still holds but we do not have
a reduction process for the $SU(N)-$observables.

In particular, we observe that, as in Section 4, a compact 
state-manifold with the
symmetry considered here gives a collapse process for $H$ and
simultaneously a reduction process for any 
one of the spins
$S$ if and only if 
it is biholomorphic to ${\bf CP}^N$ with a metric of positive
holomorphic bisectional curvature.\\

\noindent {\bf Case 3: semi-infinite with a nut} and {\bf Case 4:
semi-infinite with a bolt} have
a single critical value for the Hamiltonian, respectively
nondegenerate and degenerate, and {\bf Case 3} has positive bisectional
holomorphic curvature while {\bf Case 4} does not. The SSE 
confines the evolution to a
linear submanifold, but we cannot speak of the L\"uders postulate in
these cases since
there is only one eigenvalue.

\section*{Acknowledgements}
The authors would like to express their gratitude to D.C. Brody and 
N.J. Hitchin for stimulating discussions. N.J. Hitchin, in particular, provided a
useful introduction to the literature on K\"ahler manifolds with
positive curvature of various kinds. LPH acknowledges the hospitality
and support of the Institute for Advanced Study, Princeton, where part
of this work was carried out.

\end{document}